\documentclass[11pt,twoside]{article}
\usepackage{graphicx,epsfig,natbib,epstopdf}
\usepackage{CS18}
\begin{document}
%
% NTA: Ignore the setcounter comment here
%\setcounter{page}{7}
%
% NTA: Update your full title here
%
\title{Forward Modeling of Synthetic EUV/SXR Emission from Solar Coronal Active Regions: Case of AR 11117}
%
% NTA: enter your author name, affiliation/address information here
%
\author{V. S. Airapetian$^{1}$ and J. Allred$^{1}$}
\affil{$^1$NASA/Goddard Space Flight Center, Greenbelt, MD 20771}
%\affil{$^2$National Optical Astronomy Observatories, P.O. Box 26732, Tucson, Arizona, USA 85726}
%\affil{$^3$Institut d'Astrophysique de Paris, 98bis bd Arago, 75014 Paris, France}
%
\begin{abstract}
Recent progress in obtaining high spatial resolution images of the solar corona in the extreme-ultraviolet (EUV) with Hinode, TRACE, SDO and recent Hi-C missions and soft X-ray (SXR) bands opened a new avenue in understanding the solar coronal heating, the major goal of solar physics. The data from EUV/SXR missions suggest that solar corona is a non-uniform environment structured into active regions (AR) represented by bundles magnetic loops heated to temperatures exceeding 5 MK. Any viable coronal heating model should be capable of reproducing EUV and SXR emission from coronal active regions well as dynamic activity. Measurements of emission measures (EM) for ARs provide clues to time dependence of the heating mechanism: static versus impulsive. While static equilibrium coronal loop models are successful in reproducing SXR emission within an AR, they cannot adequately predict the bright EUV loops. Meantime, impulsive heating is capable in reproducing both EUV and SXR loop emission. The major goal of this paper is to construct realistic synthetic EM images of specific solar corona active region, AR 11117 by using our 1D fully non-linear time-dependent single-fluid hydrodynamic code. We first construct a magnetic skeleton for the entire active region using the HMI/SDO magnetogram for AR 11117 and populate magnetic field lines with plasma. We then parametrically specify impulsive heating of individual strands (flux tubes) comprising coronal loops. Next, we simulated the response of the entire active region (with LOS projection effects) to the heating function (volumetric heating rate) scaled with magnetic field and spatial scale parameters and find the best match between synthetic and actual (reconstructed) DEMs obtained by SDO.
\end{abstract}

\vspace{.2in}
\normalsize

\normalsize

%\section*{Scientific/Technical/Management}
\section{Introduction}
%\noindent
All stars later than F5 possess convective zones that drive hot corona heated to 1-10 MK. For this standpoint, the Sun has a moderately heated corona (1-3 MK) extending from the transition zone to a few solar radii. The solar coronal heating is observed in the soft X-ray (SXR) and EUV bands and plays a critical role in controlling the thermodynamics and chemistry of the Earth's upper atmosphere (Meier 1991). The corona’s variable radiative output is associated with flares and coronal mass ejections that affect space weather, and eventually, life on Earth. Variations in the radiation affect radio signal propagation and satellite drag thereby impacting communication, navigation, surveillance, and space debris collision avoidance. Predicting the spectral irradiance from the global Sun is therefore a major goal of the national space weather program. Having this capability requires an understanding of the puzzling physical mechanism that heats the outermost part of the solar atmosphere, the solar corona, to multi-million degree temperatures. Stellar SXR observations have revealed that the coronal heating processes are not unique to the Sun, but are common in magnetically active stars. Therefore, understanding the origin of high-temperature plasma in the solar/stellar coronal environments is one of the fundamental problems of solar physics and stellar astrophysics. While stellar observations show a large variety of coronal environments characterized by up to four orders of magnitude larger heating rates (for example on RS CVn stars or coronal giants), higher spatial and spectral resolution EUV/SXR observations of the solar corona provide the critical data for resolving this puzzle. Specifically, first SXR Yohkoh and later SOHO observations of the global Sun have revealed that the solar coronas represent a highly inhomogeneous environment filled with plasma frozen to magnetic structures of two basic configurations: open and closed. Magnetically open structures extend from the solar photospheres into the heliosphere, while closed structures are signified as loop-like structures filled with relatively dense (10$^9$ cm$^{-3}$) and hot (few MK) plasma emitting in EUV lines of highly ionized metals. While the quite-Sun regions are associated with weak magnetic fields (a few Gauss), EUV/SXR emitting plasma in active regions (AR) is formed in magnetic structures that can be traced back to strong (over 1 kG) surface magnetic fields. The strongest magnetic field in ARs is usually associated with hotter ($>$ 5 MK) and denser plasma which is observed as higher contrast in AIA and SXR images, while regions with weaker fields show signatures of cooler plasma. This association clearly relates the problem of coronal heating to the energy stored and released in the solar coronal magnetic field.

Energy into the magnetic field is likely supplied from the mechanical energy of photospheric convective motions.
The coronal loops observed in the AR core are usually shorter, denser with higher temperature and associated with stronger magnetic fields. The footpoints of core loops are observed in EUV structures called "moss" (Fletcher \& De Pontieu 1999; De Pontieu et al. 2013). Studies of the temperature evolution of AR coronal loops in time suggested that their emission in EUV results from impulsive heating events occurring at sub-resolution scale (or strands) and ignited a new heating scenario of coronal loops through "nanoflare storms" (Klimchuk 2006). The recent evidence in favor of impulsive heating in coronal loops comes from observations of time-lag of peaks of emission observed in high-temperature lines compared to cooler lines suggesting that these loops can be explained by so-called long nanoflare storms occurring in many strands within a coronal loop (Klimchuk 2009; Viall \& Klimchuk 2012). Recent high spatial resolution SDO and the latest High-resolution Coronal Imager (Hi-C) observations of one active region imply that a magnetic loop is not a monolithic structure, but consists of many (possibly hundreds) of unresolved "strands," with the fundamental flux tubes thinner than 15 km (Peter et al. 2013; Brooks et al. 2013). Moreover, a nanoflare scenario was further specified from analysis of cool, dense and dynamic loops observed by Hi-C observations in lower parts of coronal loops (Winebarger et al. 2013).

Two leading theories provide an explanation for how “nanoflares” release magnetic energy in the corona. Magnetic energy dissipated in coronal loops is supplied by the photospheric convection either in the form of upward propagated MHD waves (Asgari-Targhi \& van Ballegooijen 2012) or formation of current sheets driven by twisting and braiding of coronal field lines forming a nanoflare storm (Parker 1988). In either of these proposed scenarios, energy can dissipated at small scales on a single "strand" (a flux tube) in a series of transient heating events. Two important questions are: What is the time scale between two successive "nanoflares" (or frequency of nanoflares) within an AR coronal loop? To what extent are waves or current sheets responsible for nanoflare heating? These two theories predict distinctive scaling laws of the heating rates with magnetic field and characteristic spatial scales of coronal loops (Mandrini et al. 2000).
	
All coronal loop models presented to date can be divided into three categories. Early models of equilibrium loops by Rosner et al. (1978) and Craig et al. (1978) suggested that loops are symmetric, semi-circular monolithic loops with uniform cross section in static equilibrium. These and later studies of individual loops were successful in explaining many signatures of SXR and EUV loops (Porter \& Klimchuk 1995; Cargil \& Priest 1980; Aschwanden \& Schriver 2002; Winebarger et al. 2003; Reep et al. 2013). This approach is useful in studying detailed response of individual loops to different heating scenarios; however, it is difficult to compare them directly to observations of active regions with collections of loops "contaminated" by selection and line of sight (LOS) effects. Another approach is to construct three dimensional MHD models of an active region that will accommodate the above mentioned effects (Lionello et al. 2005; Gudiksen \& Nordlund 2005; Bourdin et al. 2013). These models are extremely useful in understanding a general geometry and dynamics of magnetic structures and can be directly compared to observations. However, they are computationally expensive especially when it comes to resolving physically important scales as well as in treating thermal conduction at small scales in individual loops. The third class of emerging models incorporates the advantages of individual loop models with geometry and LOS effects. This class includes forward models of active regions (Lundquist, Fisher \& McTiernan 2008a; 2008b, Patsourakos \& Klimchuk 2008; Airapetian \& Klimchuk 2009). Airapetian \& Klimchuk (2009) have developed a new class of impulsive coronal heating models that are based on introducing magnetic field extrapolation of active regions using HMI/SDO magnetograms. They make use of the "0D" HD code, EBTEL, which provides a computationally fast way to derive loop averaged temperature and density and construct 2D synthetic images of an active region driven by nanoflare storms. However, that model assumed a uniform cross section of modeled loops as well as uniform heating along each loop.

	In the current paper, we have significantly expanded on the capabilities of forward models of active regions to construct realistic synthetic images of individual ARs and the global Sun by applying our state-of-the-art fully non-linear 1D hydrodynamic code. First, we developed a fundamentally new class of active region models based on parametrically specified impulsive heating of individual strands (flux tubes) comprising coronal loops. We begin with constructing a "magnetic skeleton" of an active region using the most sophisticated methods to extrapolate Non-Linear Force Free coronal magnetic fields (NLFFF) from high resolution vector HMI/SDO and SOLIS observations (Tadesse et al 2013). We then study how the entire active region (with LOS projection effects) responds to the heating function (volumetric heating rate) scaled with magnetic field and spatial scale parameters and find the best match between synthetic and actual (reconstructed) DEMs obtained by SDO.

\section{The Magnetic Skeleton of The Solar Coronal Active Region, AR 11117 }

In this paper we construct synthetic EM images of specific ARs in EUV and SXR bands, we need first to construct a 3D equilibrium magnetic loop model of an entire AR or the "magnetic skeleton" of an active region. The magnetic skeleton in the solar corona can be realistically constructed by using SDO/HMI vector magnetograms and extrapolating them into the inner solar corona. Reliable magnetic field measurements are still restricted to the level of the photosphere, where the inverse Zeeman effect in Fraunhofer lines is observable. As an alternative to measurements in these super-photospheric layers, we must rely on numerical computations (known as extrapolation) (Amari et al., 2006) of the field that use the observed photospheric magnetic field vector as a boundary condition. These numerical computations can be carried out using potential field, force-free field or magneto-hydrodynamics (MHD) models. Force-free models do include electric current, and so they can include free-magnetic energy. Force-free models make the simplifying assumption that these currents are field aligned. A force-free model gives static representations of the state of the solar corona at a given instant. This is a good approximation in the low-β corona because the vanishing Lorentz-force does not allow currents perpendicular to the magnetic field. By applying a force-free model to a time sequence of magnetograms we can study the changes in magnetic configuration that results from a flare or eruption.

\begin{figure}[h!]
\includegraphics*[width=\linewidth]{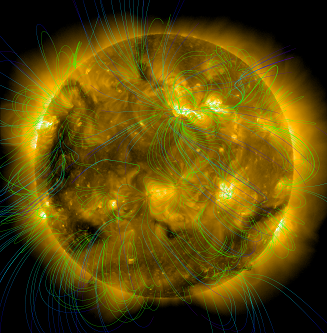}
\caption{Tracing of the magnetic field lines for the global Sun using our NLFFL extrapolation algorithm (Tadesse et al. 2013).}
\end{figure}

In nonlinear force-free field (NLFFF) models, there are no forces in the plasma which can effectively balance the Lorentz force, $\vec{J} \times \vec{B}$, (where $\vec{J}$ and $\vec{B}$ have the standard definitions of current density and magnetic field, respectively). NLFFF extrapolation is a realistic way to model the non-potential coronal fields in active regions.

\begin{figure}[h!]
\includegraphics*[width=\linewidth]{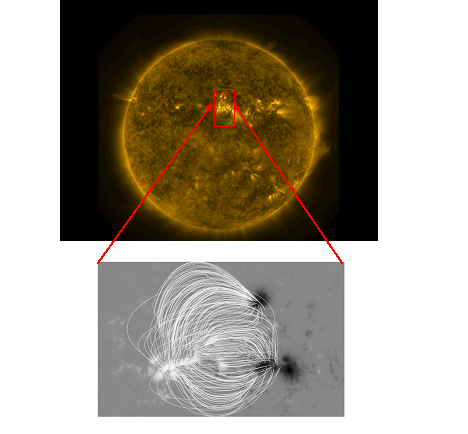}
\caption{SDO image of AR 11117 and its magnetic skeleton used for populating individual strands with plasma.}
\end{figure}

We use an optimization procedure to calculate 3-D magnetic field solutions into the corona from photospheric boundary. We implement Cartesian or spherical geometry depending on the size of area of region of interest. We have developed IDL tools which help us trace the magnetic field.

To describe the equilibrium structure of the static coronal magnetic field, the force-free assumption is appropriate:

\begin{equation}
\vec{\nabla} \times \vec{B} = \alpha \vec{B}
\end{equation}

\begin{equation}
\vec{\nabla} \cdot \vec{B} = 0
\end{equation}
			
subject to the boundary condition $B = B_{obs}$ on photosphere where $B$ is the magnetic field and $B_{obs}$ is measured vector field on the photosphere. Using the three components of B as a boundary condition requires consistent magnetograms, as outlined in Aly (1989). The photospheric vector magnetograms, obtained by the Synoptic Optical Long-term Investigations of the Sun survey (SOLIS)/Vector Spectromagnetograph (VSM) or HMI/SDO are used as the boundary conditions. Meanwhile, those measured data are inconsistent with the above force-free assumption. Therefore, one has to apply some transformations to these data before nonlinear force-free extrapolation codes can be applied. This procedure is known as preprocessing. This preprocessing scheme modifies the boundary data so that they are consistent with necessary conditions for a force-free field, namely so that integrals representing the net force and torque on the coronal volume above the photosphere are closer to zero (Wiegelmann et al. 2006; Tadesse et al. 2009). We solve the force-free equations using an optimization principle (Wheatland et al. 2000; Wiegelmann 2004) in spherical geometry (Wiegelmann 2007; Tadesse et al. 2009, 2012; 2013).

For our test calculations, we have selected AR 11117 observed by SDO on Oct 26, 2010 at 04:00 UT. The image in 171 Å is presented in Figure 2. Using the described technique, we have constructed a "magnetic skeleton" of the active region containing over 12,000 strands. We then imposed a background heating rate in each strand and evolve them using time-dependent hydrodynamics until they have reached equilibrium. Coronal loops are be treated as bundles of magnetic field lines (or elementary flux tubes) that expand into the corona but are rooted in the solar photosphere. Their lengths are much greater than their widths and their orientation is along the direction of the magnetic field. The expansion factor (cross-section) of each individual flux tube is controlled by the condition of the magnetic flux conservation along the tube, specified at the photosphere from the local magnetic field derived from a magnetogram and the minimum size of the magnetic element resolved by HMI observations, ~350 km.

\section{ARC7: 1-D Hydrodynamic Model of the AR 11117}

Once the magnetic skeleton of the active regions is constructed, we populate each strand of the active region with an initial atmospheric state. To do this we apply uniform background heating, $E_{bg}$, that provides the temperature of 0.5 MK. This allows density and temperature to reach a steady-state equilibrium. To simulate the thermodynamics in each coronal loop driven by a storm of impulsive "nanoflare" events, we use a time-dependent heating rate applied to each strand. The heating rate we use has a general form allowing us to model energy release due to a number of different physical mechanisms. The time-dependence of the impulsive heating from each nanoflare were modeled as triangular pulse with a maximum value given by,

\begin{equation}
\centering
E_H = E_{bg} + g(t) E_0
\end{equation}

After heating has been applied for a specified duration we continue to simulate the strands as they cool. The duration of each pulse as well the number of heating pulses applied, and the cooling time are also free parameters. Varying these allows to study the frequency with which nanoflare heating occurs.
Therefore, the physical size of the cell varies for each loop with its length with the grid resolution of a few tens of km at the loop base. The heating function, $E_0$, is scaled with the local value of the magnetic field within each {\it nth} cell as $B_{n}^{\alpha}$ and the physical extent of the cell as $l_{n}^{\beta}$. Therefore, in each cell the local heating function is defined as

\begin{equation}
\centering
E_0=\epsilon~B_{n}^{\alpha}~\Delta{s}_{n}^{\beta}
\end{equation}

In the low solar corona, the magnetic forces dominate over gas pressure. In this regime, plasma is constrained to flow along magnetic field lines and the magnetic field remains static over the time scales which we simulated. Thus, the full 3D MHD equations can be well-approximated by one-dimensional hydrodynamics with that dimension being the axis of magnetic field lines. To model solar coronal loop dynamics, we solve the 1D hydrodynamic equations using a modified form of the ARC7 code (Allred \& MacNeice 2012). ARC7 was created to solve the equations of MHD in 2.5D geometry. It solves the equations of MHD explicitly using a 2nd order accurate in time and space flux-corrected transport algorithm. A radiative loss term is included in the energy conservation equation. This term is proportional to $n_{e}^2~{\Lambda}(T)$, where $n_e$ is the electron number density and ${\Lambda}(T)$ is the radiative loss function and is obtained from the CHIANTI package (Dere et al. 2009).

Field-aligned thermal conduction is included in the energy conservation equation and is assumed to have the classical Spitzer formulation. However, during our impulsive heating simulations temperature gradients can occasionally become large enough that the Spitzer formula predicts fluxes which would exceed the free electron streaming rate. This is unphysically large and we cap the heat flux at the free streaming rate. In order to capture the effect of the expansion of the magnetic field from the footpoints into the corona, we scale the cross sectional area of ARC7’s grid cells so that magnetic flux is conserved. At the loops boundaries (i.e., footpoints) we have implemented a non-reflecting boundary condition so that waves can pass through. The boundary of our loops is held at a temperature of 20,000 K and start with sufficient mass density so that material can be evaporated into the corona in response to impulsive heating without significantly changing the boundary density.

We perform an impulsive heating simulation using the following algorithm. An initial background heating rate is specified to obtain an equilibrium temperature ~0.5MK.  We use the RTV scaling laws (Rosner et al. 1978) to setup a starting atmospheric state within loops depending on the background heating rate and loop length. We allow ARC7 to evolve the loop until it reaches equilibrium. We then turn on the impulsive heating term which linearly ramps the heat function up until it reaches a maximum value and then linearly ramps it down over a time $\delta$t. The maximum value heating function is assumed to have the form $Q_0~ B^{\alpha}/{\Delta}s^{\beta}$, where $Q_0$ is a coefficient, $B$ is the magnetic field strength and $\Delta$s is the length of the element along a flux tube. We also specify n, ${\Delta}t_{int}$, and ${|Delta}t_{cool}$, where  n=4  is the number of heating pulses we applied during the simulation, ${\Delta}t_{int}$  is the time interval between heating pulses and ${\Delta}t_{cool}$ is the time we allow the loop to cool after the impulsive heating has been applied .

We have chosen to use ARC7 because of its high-speed performance. As noted, ARC7 is a 2.5D MHD code. Our proposed method requires hydrodynamics in only one spatial dimension because plasma is frozen-in the magnetic field in a low-$\beta$ low corona  We have simplified ARC7 to take advantage of these assumptions which results in a vast improvement in performance. Using a standard single processor computer, we can model the evolution of a single loop in response to impulsive heating in a few seconds. Our model active regions have on the order of 104 individual strands. We performed these strand simulations in parallel using 100 processors simultaneously on NASA’s Pleiades supercomputer and completed the simulations over an entire active region in about an hour.

To reproduce the magnetic structure of the active region, we have used 12,800 individual strands and ran individual trains of nanoflares (low frequency events) on each of them. We then ran nanoflare trains on each of the strands. We selected the duration of a nanoflare as $\Delta$ t = 200 s with the time interval between two successive events, $\tau$ = 200 s. For this example we have used $\alpha$ = 2 and $\beta$ = 2.

\begin{figure}[h!]
\includegraphics*[width=\linewidth]{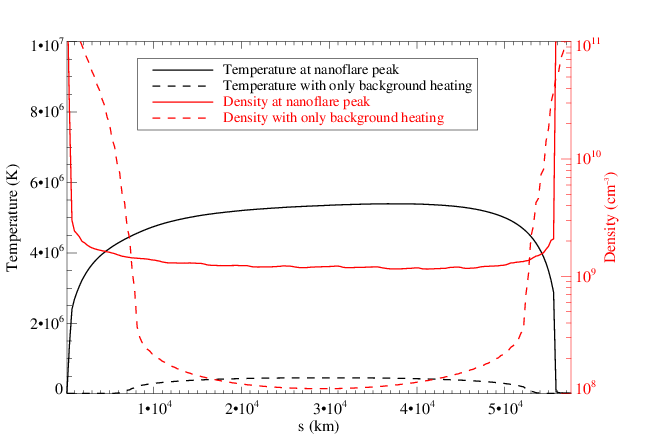}
\caption{Distribution of of plasma temperature and density along a typical loop with the length of 58 Mm during a train of 5 nanoflares within one flux tube of the coronal loop for the case of background heating only (dashed line) and for the case of nanoflare heating (solid line) at its peak.}
\end{figure}

\noindent
In the selected extrapolation model of the active region, the loop lengths vary between 5Mm and 200 Mm. We ran simulations with 5 consecutive pulses then another 5000 s of cooling time. Figure 3 shows the temperature and density at the flare peak in a single strand compared with the background temperature and density. The temporal evolution of the peak of that strand is shown in Figure 4.

\begin{figure}[h!]
\includegraphics*[width=\linewidth]{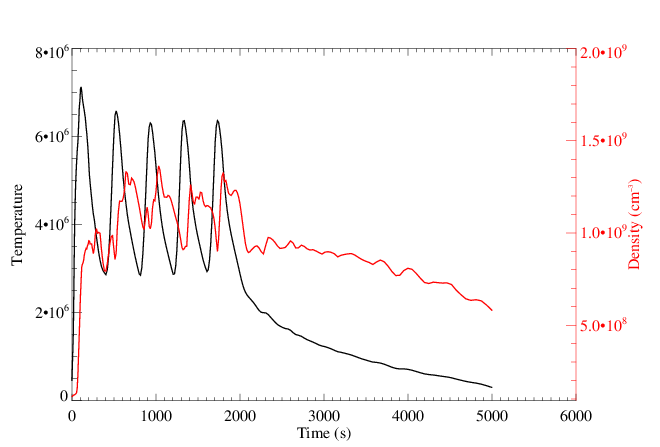}
\caption{Distribution of of plasma temperature and density along a typical loop with the length of 58 Mm during a train of 5 nanoflares within one flux tube of the coronal loop for the case of background heating only (dashed line) and for the case of nanoflare heating (solid line) at its peak.}
\end{figure}

\section{Synthetic DEM Images of the AR 11117}.

We combined the results of all of our HD simulations to form a 2D picture of the DEM for that active region. We calculated the DEM in each grid cell of each strand using our 1D simulations. We then averaged these DEM in time over the duration of the simulations. Time-averaging captures the assumption that these impulsive heating events occur at random intervals and are independent of each other. These time-averaged DEMs were projected along the line-of-sight back onto HMI pixels forming a 2D representation of the temperature and density structure of that AR. Once the DEM is known, we calculated the optically-thin radiation spectrum, I(${\lambda}$), using the most recent version of CHIANTI atomic database package (Dere et al. 2009).

\begin{figure}[h!]
\centering
\includegraphics*[width=\linewidth]{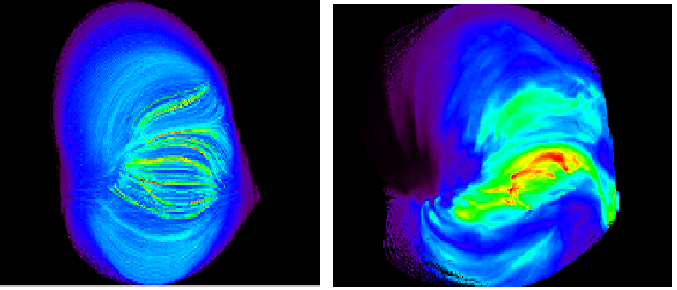}
\caption{Left panel - Time-averaged DEM for AR 11117  from our 1D HD model are projected along the line-of-sight back onto HMI pixels. Right panel-DEM for AR 11117 reconstructed from AIA images (see the text).}
\end{figure}

The right panel of Figure 5 shows an example of the DEM constructed from our simulations. Our model results can be compared with observations in two ways. First, we can convolve our DEM with AIA filter passbands to produce synthetic images which can be compared directly with AIA images. We can also construct a DEM from AIA images and compare that directly with our simulated DEM. Developing methods for constructing DEMs from AIA images is a very active topic of research. We have used the tool developed by Hannah \& Kontar (2012). This tool uses a regularized inversion method and has the advantage that it provides uncertainties in both the DEM and temperature (i.e., it provides both horizontal and vertical error estimates). The AIA images were obtained and processed using SolarSoftWare (SSW) IDL packages. We downloaded level 1 AIA images for all passbands for the time interval over which our HMI magnetogram was observed using the SSW routines vso$\textunderscore$search and vso$\textunderscore$get. Next, we converted them to level 1.5 and co-aligned them with the HMI magnetogram using the aia$\textunderscore$prep function. Finally, we ran the DEM construction program data2dem$\textunderscore$reg provided by Hannah \& Kontar (2012) on all pixels which modeled in our simulations. The left panel of Figure 5 shows this DEM reconstruction at a temperature of log T = 6.5. This simulated DEM distribution will be compared with the observationally derived DEM for the active region in the near future.

\begin{figure}[h!]
\centering
\includegraphics*[width=\linewidth]{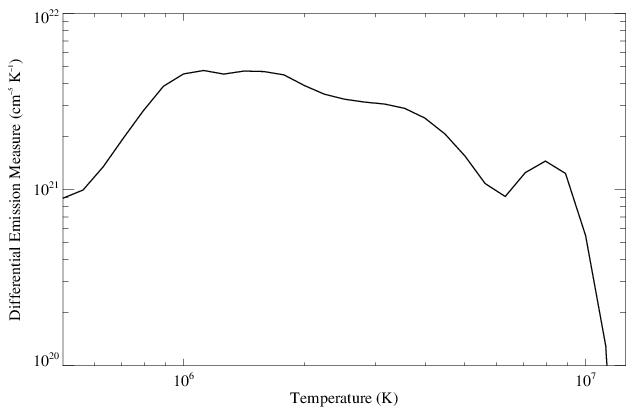}
\caption{Simulated Differential Emission Measure distribution for the entire coronal active region AR 11117.}
\end{figure}

\section{Conclusions}

We have constructed the first realistic synthetic EM images of the entire coronal active region, AR 11117 driven by a storm of nanoflares. Each nanoflare event was modeled by using our 1D fully non-linear time-dependent single-fluid hydrodynamic code. We simulated the response of the entire active region to a storm of nanoflares specified by impulsive (time-dependent) heating function occurring on over 12,000 strands within the active region. The heating function is scaled with the magnetic field and spatial scale parameters with $\alpha$=2, $\beta$=2 power indices. The reconstructed DEM for this AR will be compared with the observationally derived DEM for the active region in the near future. We will also construct DEMs for a range of $\alpha$ and $\beta$ values to determine the sensitivity of its shape to the specified shape of the heating function.

{}

\end{document}